\documentclass[11pt,preprint,graphicx]{aastex}

\begin{document}

\title{Network Tools for Astronomical Data Retrieval}

\author{James Schombert}
\affil{Department of Physics, University of Oregon, Eugene, OR 97403;
js@abyss.uoregon.edu}

\begin{abstract}

The first step in a science project is the acquisition and understanding of
the relevant data.  This paper outlines the results of a project to design
and test network tools specifically oriented at retrieving astronomical
data.  The tools range from simple data transfer methods to more complex
browser-emulating scripts.  When integrated with a defined sample or
catalog, these scripts provide seamless techniques to retrieve and store
data of varying types.  Examples are given on how these tools can be used
to leapfrog from website to website to acquire multi-wavelength datasets.
This project demonstrates the capability to use multiple data websites, in
conjunction, to perform the type of calculations once reserved for on-site
datasets.

\end{abstract}


\section{INTRODUCTION}

\noindent In the philosophy of science there is espoused a view that
science moves in revolutions (Kuhn 1962), abrupt changes in the framework
of scientific understanding in particular fields.  Historical examples are
global theories such as atomism and relativity.  Under this concept of how
science is done, between revolutions (or paradigm shifts) researchers are
involved in 'puzzle-solving' type of science (normal science) attempting to
stretch the limits of the current paradigm.  New extremes pose problems
for the current paradigms and leads to the next shift.

\noindent However, this view of science ignores a critical component to
science, discovery.  It is fair to say that most of our critical ideas in
astronomy over the past few decades were not due to a paradigm shift or
puzzle-solving science but rather due to discovery (e.g., dark matter, dark
energy, quasars, Butcher-Oemler effect, etc.).   It's also becoming obvious
that few of our theoretical ideas (computer models and
simulations) are relevant beyond a few decades, but our discoveries last
forever.

\noindent We (the astronomical community) are entering a new era of
discovery science with the advent of all-sky, multi-wavelength,
spectrophotometric and imaging archives.  Later generations will look back
on this time as a golden age for astronomy where new technologies and space
missions opened regions of the electromagnetic spectrum previously
unexplored.  In fact, the current primary inhibitor to discovery science is
our ability to search, sort and analyze our datasets.

\noindent During the 20th century, solid scientific progress was led by a
combination of new technologies plus the computational power to analyze
the output from these new technologies.  Numerous are the published papers
where a discovery hinged on some software tool or computational method
to interpret the data.  In addition, the turn of the century saw a sharp
change from proprietary datasets to public domain, widely distributed
datasets, and the new paradigm where one's ability to search, query and
understand the growing datasets defines the science to be done.

\noindent For many projects, the successful achievement of their science
goals depends on their ability to perform {\it e-science}, the capability to
gather and analysis the appropriate data.  To this end, network tools,
tools that enhance a researchers ability to gather data from networked sites,
have become an increasing important weapon in a researchers arsenal.
Discoveries by exploration of parameter space first require the samples be
defined, then acquired.  The new breed of researcher understands datasets,
and how to gather the data.

\section{Network Tools}

\noindent All the tools developed for this project use the Python scripting
language (www.python.org).  Python has the advantage of being 1) easy to
learn, 2) available on all operating systems (thereby, any scripts you
write are easily transportable from system to system) and 3) contains
numerous modules designed specifically to handle network issues.  A
scripting language also has the advantage of lacking a compiler, thus, it
is easy to operate and flexible to work with.  The Python language also has
the unique characteristics of being well designed to work with text based
data as well as numerical data, handles files and directories with little
effort and enjoys an excellent try/except failure mode which is robust from
errors that may cause a script to crash.

\noindent The reader is assumed to be mildly familiar with the Python
language for this article is not a tutorial in Python usage (there are many
on the web).  However, the typical user will have no trouble building on the
many examples provided by this project, while at the same time advancing
their knowledge of the Python language.  In addition, several
Python software projects are found within the astronomical community (e.g.
PyFITS, PyIRAF, Numpy, SciPy), and there is a growing number of modules to
work with data, images, GUI's, etc.  Thus, there are many avenues for a
researcher to quickly jump in and starting coding.

\noindent Much of the work described herein is an offshoot of the author's
ARCHANGEL galaxy photometry system (abyss.uoregon.edu/~js/archangel).  That
system found it useful to 1) pull the imaging data from some website
archive (e.g., HST, 2MASS, DSS), 2) analysis the raw image and backup the
results up to a separate system and 3) retrieve from NED (NASA's
Extragalactic Database) all the relevant information for calibration (e.g.,
galactic extinction, redshift, etc.).  Each of these steps required the
use of some script that could access data across the network and/or
transfer files from place to place.

\noindent In the interest of generating a new network tools community, all
the scripts discussed in this paper are available for download from
abyss.uoregon.edu/~js/network.  I sincerely hope that the reader
finds them useful, learns, modifies them and, most importantly, sends me
(jschombe@uoregon.edu) feedback on what works (and what broke) as well as
ideas for future tools and research direction.  This follows the model of a
virtual community for computing knowledge.

\section{Data Transfer by SSH/SFTP}

\noindent Perhaps the most common method of file transfer and retrieval is
sftp (secure shell file transfer protocol) using some version of OpenSSH
protocols.  This is certainly the most popular method of transferring data
from one friendly system to another (friendly meaning you control both
systems; e.g., transfer of data from the telescope to your office
computer).   A script that uses sftp would take the place of time consuming
command line typing and, using Python's ability to search and parse file
directories.  Even a small script eases the laborious typing required to
transfer a mixture of file types.

\noindent The simplest manner to communicate to the shell from Python is
the use of the {\tt os.system} command (although the much more complicated
{\tt subprocess} module is now recommended).  The following example pushes
a bunch of FITS files onto a remote system by making a temporary file with
sftp commands, then uses {\tt os.system} to send a 'sftp -b' commands to
the shell.

\begin{verbatim}
import os
file=open('sftp.cmds','w')
file.write('cd /data')
file.write('put *.fits')
file.write('quit')
file.close()
os.system('sftp -b sftp.cmds user@some_data_site')
os.system('rm sftp.cmds')
\end{verbatim}

\noindent This type of operation works well as a cron job for moving backup
files during off hours, or any background task that doesn't require human
monitoring.  However, this type of script is clumsy (e.g., requires the
creation and destruction of the temporary file, {\tt sftp.cmds}).  A
smoother interface uses the {\tt pexpect} module
(sourceforge.net/projects/pexpect), as in the following example:

\begin{verbatim}
import pexpect
p=pexpect.spawn('sftp user@some_data_site.edu')
p.expect('sftp>')
p.sendline('cd /data')
p.sendline('put *.fits')
p.sendline('quit')
\end{verbatim}

\noindent Also note that both scripts require a connection with the host
that uses a 'no password' public key (easy to step up for machines you
own).  If the script requires a password, you will be prompted for it
(thus, failing for a cron job).  Automatic passwords are more difficult in
scripts as they must come from pty rather than stdin.  And that information
is too powerful for this paper.

\noindent Simple transfers seem the least necessary to automate.  If one
controls both systems, the transfers are not time critical.  However, one
can imagine scenarios where scripts of this type may be useful.  For
example, one could run a script in background, say every 5 minutes, that
identifies recent data, and ships it down range to avoid loses from
catastrophic failures at one end (e.g., telescope computer hard drive
failure).

\section{Data Transfer by URL's/HTTP}

\noindent By the 1990's, the standard method of distributing data was
through the use of websites.  In fact, for a majority of projects mandated
to distribute data, a webpage is the fastest technique to comply to the
requirements.  A remote system interacts with websites through the use of
URL's (uniform resource locator). Access through URL's is the
responsibility of the {\tt urllib} module in Python.  This module allows a
script to send a request to a website, read the return HTML file and store
in memory.  A simple example is the following:

\begin{verbatim}
import urllib
page=urllib.urlopen('http://a_webpage.com').read()
\end{verbatim}

\noindent Of course, the returning data is the HTML that makes up the
webpage, which is usually not the most transparent format for extracting
data.  Parsing the HTML to extract a value can be tricky, although there
are modules for extracting tabular data (e.g. BeautifulSoup,
www.crummy.com/software/BeautifulSoup).  A simple command to strip all the
HTML commands uses the regular expression ({\tt re}) module (i.e., {\tt
re.sub('<.*?>','',page)}).  This will leave you with all the words and
numbers outside the hypertext tags.  It's also possible to identify
specific pieces of information in a webpage.  For example, a favorite comic
strip image by searching on "src img=" tag, then striping the identifier
tag.

\noindent Again, using an urllib script as a cron job allows a user to monitor
websites for changes or new data.  The user can be alerted by email using
the {\tt smtplib} module, where the script can email a message through an
approved SMTP server (see example below).  This is particular useful for time critical
information (sudden change in your bank account? opening in a class you
want to attend?).

\begin{verbatim}
import smtplib
server=smtplib.SMTP('smtp.gmail.com')
msg='Content-Type: text/html\nSubject: Automatic Email\n\n<html><pre>\nA message!'
server.sendmail('mail_bot@your_machine','user@gmail.com',msg)
server.quit()
\end{verbatim}

\noindent Some websites maintain a consistency to their HTML format such
that quick information can be extracted with a simple script.  For example,
the following script grabs the J2000 coords for a galaxy from NED:

\begin{verbatim}
import urllib, sys
name=''.join(sys.argv[1:])
page=urllib.urlopen('http://nedwww.ipac.caltech.edu/cgi-bin/nph-objsearch?'+ \
                    'objname='+name+&extend=no&out_csys=Equatorial'+ \
                    '&out_equinox=J2000.0&obj_sort=RA+or+Longitude'+ \
                    '&of=pre_text&zv_breaker=30000.0&list_limit=5'+ \
                    '&img_stamp=YES').read()
for t in page.split('\n'):
  if 'Equatorial' in t and 'J2000' in t:
    print ' | '.join(t.split()[:6]),'|'
    break
else:
  print 'object not found in NED'
\end{verbatim}

\noindent Note that the object name is all the words after the command
(e.g. {\tt ./ned.py NGC 4881}).  The parsing is done by NED, the webpage is
piped back to the script.  The script then splits by carriage returns
looking for the line that has the coordinates.  The secret here is that NED
always maintains the same 'look', and the coordinates are always on the
line with unique identifiers 'Equatorial' and 'J2000'.

\noindent Again, this is not an elegant method to communicate with a data
archive, but it is the simplest.  Some investment in time is spent decoding the
source hypertext of the webpage to find the particular set of lines from
which to extract the values.  Thus, this method is hardly efficient if one
needs more information than a simple set of coordinates.

\noindent To capture the full collection of data on a galaxy, NED offers an
XML output to their queries.  In order to access the XML file, one simply
changes the URL by adding "of=xml\_all".  This returns the entire set of
NED data on the query galaxy in an easy to parse XML format.  NED also
offers several XML files for photometry data, reference data, etc (see the
tools website for a suite of NED scripts).  To work with the returning XML
file, Python has a number of XML modules.  However, this project has
constructed one that better matches astronomical data and is discussed in
the next section.

\section{XML processing}

\noindent Storage of data in XML format closely mimics the HTML format that
make up webpages through the use of tags to identify each data element.
Each element (or data atom) has attributes, data and children associated
with it.  For example, an element 'redshift' may have the data value of
35,444 and an attribute of 'units=km/sec'.  Children are addition elements
embedded inside the parent element.  XML files are not particularly
readable, but as they are stored in raw ASCII format and are, therefore,
very transportable.

\noindent To ease the conversion of data into XML format (and its
extraction), this project offers an XML module ({\tt xml\_archangel}) based
on Python's xml.dom routines.  This module was designed for storage of
galaxy photometry data; however, is flexible to accommodate any type of
data as well as arrays.  The module offers two basic classes, xml\_read and
xml\_write.  The xml\_read class takes a standard XML file and parses into
a Python list of elements using the following commands:

\begin{verbatim}
from xml_archangel import *
doc = minidom.parse(file)
rootNode = doc.documentElement
elements=xml_read(rootNode).walk(rootNode)
\end{verbatim}

\noindent The resulting list, {\tt elements} is packaged into three parts, its
attributes (as a dictionary), its data and its children (also as a
dictionary).  Thus, each element appears in the script as the following:

\begin{verbatim}
                [{attributes},data,{children}]
\end{verbatim}

\noindent Using the standard Python notation, the attributes and children of
an element are in the form of a dictionary, the data are a unicode string.
The children elements, of course, are stored in the same structure, which
allows recursive searching for nested elements.  Note that this element
list can be modified by the script, then output to a file with the
xml\_write call.  The xml\_archangel script allows the user to pull or push
elements into an XML file, add arrays or print a tree of the entire
list.

\noindent Arrays are handled in a slightly different fashion.  Following
the recommendations of the VOTable project, arrays are stored as element
name 'array' (attributes that indicate the name of the array) with each
array having N children called 'axis'.  For example, an array of sky box
positions:

\begin{verbatim}
<array name='sky_boxes'>
  <axis name='x'>
    45
    65
    33
  </axis>
  <axis name='y'>
    23
    55
    11
  </axis>
  <axis name='size'>
    20x20
    20x20
    10x10
  </axis>
</array>
\end{verbatim}

\noindent While this is not the most readable format, it is easy to parse
in a Python script.  The script can then convert this into a numerical
array for processing with the extremely useful numpy routines
(http://numpy.scipy.org) that bring all the power of a C++ processing
routine into Python.

\section{Image Extraction (DSS/2MASS)}

\noindent Most of the common data archives use a simple POST/GET webpage to
access their data using HTML FORM methods.  One example is the DSS archive
(archive.stsci.edu/cgi-bin/dss\_form) where the user enters a name or
coordinate of interest and selects the type of Palomar Sky Survey image to
be downloaded.  Standard HTML FORM stores the user selected variables (in
the source webpage as <input> tags) and passes them into a new URL with the
variables in the format of "\&variable=value".  So the user can use the
webpage to enter the variables or, if they know the variable names, they
can simply type the URL themselves.

\noindent Any website that uses a HTML FORM can be parsed into a URL for a
Python script.  Some detective work is needed, for example, searching
through the source to identify all the variables.  Or the user can make a
simple search, then copy/paste the URL from the navigation bar on their
browser into the Python script (noted what variables control the object).
There are also tools available in the common browsers to diagnose a webpage
(e.g., the Web Developer add-on in Firefox).

\noindent The examples at our project website list two scripts (too long
for this article, although only 50 lines in total length), one to access
DSS images from STScI and the other to extract images from 2MASS.  Both
take advantage of NED's website to find a galaxy's coordinates, parse a URL
using those coordinates (and user selected field size, image type, etc.)
then upload that URL to DSS/2MASS ("squirt the bird" in NASA terminology).
The script then reads the return data stream and writes out a FITS file.
Slight modification to the script allows images to be stitched together, or
multiple bandpasses to be built into a hypercube.

\noindent This is a good point in our discussion to mention abuse.  These
scripts are indented for use on small samples (less than 50 or so).
Downloading, for example, the entire UGC catalog from DSS is not an
efficient use of network time.  For extremely large samples, the user is
encouraged to contact the project in question for extraction of the needed
data on-site.  The projects are always helpful working with large projects,
and collaboration with the projects for this type of research sends a
strong message to the funding agencies.  Bottom line, use some common sense
in the amount of data you are requesting from websites intended only for
the exchange of a few images at a time.

\section{Cookies and Passwords}

\noindent More sophisticated websites require passwords and use cookies to
prevent a user from spoofing the URL directly to the data.  Python also has
the ability to store cookies in an automatic fashion using the {\tt
httplib} module.   A typical interaction with a website with a password and
session cookie would look like the following:

\begin{verbatim}
import httplib, urllib
userid='joe_user'
password='a_password'
urlencoded = urllib.urlencode({'user': userid, 'pin': password})
hlink = httplib.HTTP('the_website.com')
hlink.putrequest('POST', '/the_area_of_interest')
hlink.putheader('Cookie', 'SESSION_ID=set')
hlink.putheader('Content-type', 'application/x-www-form-urlencoded')
hlink.putheader('Content-length', '%d' % len(urlencoded))
hlink.endheaders()
hlink.send(urlencoded)
errcode, errmsg, header = hlink.getreply()
if errcode == 503:
  print 'website off-line'
  sys.exit()
mark=str(header).index('Cookie')
cookie=str(header)[mark+15:mark+31]
page=hlink.getfile().read()
\end{verbatim}

\noindent In this example, every further page request requires a new {\tt
hlink.putrequest('GET','\\new\_place')} and a new cookie is sent by the
website to be tested for the next page request.  Note that the user must
have a legitimate ID and password, this is not a technique to hack a
website and it is assumed the user has authorized access to the data.

\noindent The uses for this type of script are endless.  Written as a cron
job, this routine can monitor your bank account, credit card, class lists
or grant proposals.  Again, when matched with the {\tt smtplib} module,
such a script becomes a powerful email alert system.

\noindent Another use for a script of this type is the automatic submission
of data.  For example, my University requires that student grades be
entered into a website using drop-down menus for each student ID.  If the
class contains 200 students, this can literally be a several hour activity.
Instead, a short script can be written to login into the website, grab a
file of student grades on the local machine and POST them to the website by
student ID (although your local network services might be curious on how
you entered 200 grades in 3.5 microseconds).  Again, this technique is open
to abuse and a responsible user would limit the number of interactions with
a website, and their frequency (e.g., placing the {\tt time.sleep(1)}
command between page requests).

\section{Behaving like a Browser}

\noindent Websites have become increasingly complex in recent years, often with
complicated cookies that the {\tt urllib} and {\tt httplib} modules fail to
handle.  However, every website must interact with a browser, so nothing
can be encoded or hidden that can't be parsed by whichever browser the user
selects.  Ultimately, the best script is one that behaves like a browser
and can be trained to proceed to the internal pages of interest (i.e.
clicking the buttons).  This is the job of the {\tt mechanize} module
(wwwsearch.sourceforge.net/mechanize).

\noindent There are numerous examples at the {\tt mechanize} website, but the
following is a simple use in a script:

\begin{verbatim}
from mechanize import Browser
from mechanize import UserAgent
  b = Browser()
  b.addheaders=[('User-Agent', 'Python script')]
  b.open('https://secure_website.com')
  userid='user_id'
  password='a_password'
  b.select_form(name='LoginForm')
  b['userID']=userid
  b.submit()
  for form in b.forms():
    print form
\end{verbatim}

\noindent Note that the website may reject User-Agent's that are not a known
browsers.  This script must be trained, in the sense that the user probably
needs to manually follow the website paths first, then copy those paths
into the script.  And more detective work is probably required on the FORM
variables and their usage.

\noindent The ultimate goal for a script that uses mechanize is to
parse and understand what a webpage means, and use that information to make
decisions.  This would form the front end of a thinking or knowledge
system, one that harvests information at a higher level than just reducing
the data from tabular form.  This will be the focus of our future work.

\section{Summary}

\noindent The goal of this paper is to outline some simple network tools to
enhance the retrieval of astronomical data from local machines or data
archive websites.  Hopefully, these scripts improve the efficiency of a
researchers to find and acquire the information they need to address their
science questions.  Less time spent managing files and directories means
more time spent on analysis and understanding.

Some examples of uses for these scripts are:

\begin{itemize}

\item{} Transfer of backup files during off hours by cron jobs
\item{} Monitor files and submit email alerts for multiple systems
\item{} Retrieve and parse a webpage
\item{} Extract a value from a webpage
\item{} Submit a request and respond from a webpage
\item{} Pull XML data from a webpage
\item{} Interact, in an automatic fashion, with a website that uses a
ID/password
\item{} Behave like a browser, parsing requests and designing responds
interactively with a website

\end{itemize}

The reader is invited to modify or add to the network library.  Simply send
your comments and scripts to jschombe@uoregon.edu and we will post them on
the growing website. 

\acknowledgements

Financial support from NASA's AISR program is gratefully acknowledged.
Many of these tools were developed to support efforts of our NASA and NSF's
data centers, and ongoing efforts for the US National Virtual Observatory,
which is sponsored by the NSF.

\end{document}